\documentclass[prl,aps,twocolumn]{revtex4}
\usepackage{amsmath,amssymb,graphicx,verbatim}

\begin{document}

\title{Composite pulses for robust universal control of singlet-triplet qubits}

\author{Xin Wang$^{1}$, Lev S.~Bishop$^{1,2}$, J.~P.~Kestner$^{1}$, Edwin Barnes$^{1}$, Kai Sun$^{1,2}$, and S.~Das Sarma$^{1,2}$}

\affiliation{$^{1}$Condensed Matter Theory Center, Department of Physics,
University of Maryland, College Park, Maryland 20742, USA\\
$^{2}$Joint Quantum Institute, University of Maryland, College Park, Maryland 20742, USA}

\date{\today}

\begin{abstract}

Precise qubit manipulation is fundamental to quantum computing, yet experimental systems generally have stray coupling between the qubit and the environment, which hinders the necessary high-precision control.
We report here the first theoretical progress in correcting an important class of errors stemming from fluctuations in the magnetic field gradient,  in the context of the singlet-triplet spin qubit in a semiconductor double quantum dot.
These errors are not amenable to correction via control techniques developed in other contexts, since here the experimenter has precise control only over the rotation rate about the $z$-axis of the Bloch sphere, and this rate is furthermore restricted to be positive and bounded. 
Despite these strong constraints, we construct simple electrical pulse sequences that, for small gradients, carry out $z$-axis rotations while canceling errors up to the sixth order in gradient fluctuations, and for large gradients, carry out arbitrary rotations while canceling the leading order error.

\end{abstract}

\maketitle

\section*{Introduction}

A quantum computer would permit exponentially faster algorithms than an ordinary computer for certain important types of problems  \cite{NielsenChuang.00}. Universal quantum computation requires the ability to perform an entangling two-qubit gate and precise single-qubit rotations around two different axes of the Bloch sphere. Single-electron spin qubits in semiconductor quantum dots
potentially have marked advantages in fast two-qubit gating and scalability \cite{Loss.98}, but suffer an embarrassing difficulty in performing fast single-qubit rotations since the strong, high-frequency magnetic fields na\"{i}vely required  \cite{Koppens.06} heat the sample and are hard to confine to a single qubit  \cite{Pioro-Ladriere.06}.  This problem is circumvented by encoding the qubit in the low-lying singlet-triplet subspace of a two-electron double quantum dot \cite{Levy.02,Petta.05, Bluhm.11, Maune.12}.  Fast electrical control of the exchange coupling, $J$, via the tilt of the effective double-well potential allows sub-ns rotations about the $z$-axis of the Bloch sphere.

In order to perform arbitrary rotations of such a qubit, though, one must introduce a difference, $\Delta B$, between the local magnetic fields at each dot, resulting in rotation about the $x$-axis of the Bloch sphere.  This can be done either by pumping a nuclear spin polarization gradient \cite{Foletti.09} or by depositing a micromagnet nearby \cite{Brunner.11}.  However, a problem remains:  The local magnetic field typically fluctuates slowly due to second-order nuclear spin flip-flops mediated by the hyperfine coupling to the electron spin \cite{Reilly.08,Cywinski.09} and due to charge-noise-induced shifts of the double dot position in the inhomogeneous field.  The resulting uncertainty in $\Delta B$ introduces a quasi-static random component to the rotation about the $x$-axis which, in the course of ensemble averaging, leads to rapid decoherence of the qubit on the free induction decay timescale of $T^{\ast}_2$.  However, the fact that these errors implement coherent rotations (albeit by an unknown angle) makes it possible to reduce their effect by means of dynamical control; this nice feature is due to the
non-Markovian nature of the nuclear spin bath, a situation unique to quantum dot spin qubits. In the case of quantum memory, dynamical decoupling techniques \cite{Carr.54,Meiboom.58,Uhrig.07,Witzel.07,Cywinski.09,Barthel.10,Bluhm.11} can be employed to preserve qubit information long beyond $T^{\ast}_2$, up to a timescale $T_2$ (which is defined with
respect to a specific dynamical decoupling sequence) at which this information is ultimately lost due to dynamical fluctuations.  This ability is crucial, since typically $T_2 \gtrsim 10^4 T^{\ast}_2$ for localized electron spins in semiconductors -- in particular, in GaAs  quantum dot systems \cite{Petta.05,Bluhm.11},
$T^{\ast}_2 \sim $ 10 ns, $T_2 \sim$ 0.1  ms, and in Si \cite{Maune.12}, $T^{\ast}_2 \sim$ 100 ns with $T_2$ predicted \cite{Witzel.10} to be $\sim1$ ms.  
However, such echo techniques cannot be performed simultaneously with arbitrary single-qubit rotations, so gate errors are still dominated by statistical fluctuations and depend on the ratio of the gate time to $T^{\ast}_2$ rather than to $T_2$. It is of utmost importance to address this problem since fault-tolerant quantum computation requires extremely precise single-qubit rotations.  Thus the task is to find dynamically corrected gates \cite{Goelman.89,Grace.11,Khodjasteh.09,Khodjasteh.10,Bensky.10} applicable to singlet-triplet qubits. 

Finding such gates is challenging because available control in real experimental singlet-triplet spin qubit systems is rather limited: One only has precise control over the rotation rate about the $z$-axis of the Bloch sphere via the exchange interaction, and due to the nature of the exchange interaction this rotation rate is intrinsically restricted to be positive and bounded. Meanwhile, the rotation around the $x$-axis due to the magnetic field gradient cannot be precisely controlled. 
These control constraints specific to singlet-triplet qubits render the numerous quantum control techniques developed in other fields, such as nuclear magnetic resonance, inapplicable. 
In this work, we show how to perform dynamically corrected single-qubit gates on singlet-triplet qubits, dramatically reducing errors while fully respecting these experimental constraints. 
We construct simple electrical pulse sequences that, for small magnetic field gradients, carry out rotations about the $z$-axis while canceling gate errors up to the sixth order in the gradient fluctuations, and for large magnetic field gradients, carry out arbitrary rotations while canceling the leading order error.
This represents an important step forward in the development of singlet-triplet qubits as viable resources for quantum computing.

\begin{figure}
\centering
\includegraphics[width=0.95\columnwidth]{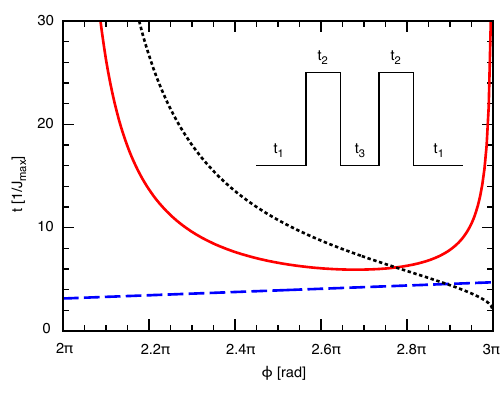}
\caption{Example of the five-piece \textsc{supcode} pulse canceling gate error up to the fourth order in $\delta h$ for $h_0=0$. The inset shows a schematic profile of such a pulse as a function of time. The time duration of each piece is denoted by $t_1$, $t_2$, and $t_3$. In the main panel we show the values of $t_1$ (red solid line), $t_2$ (blue dashed line), and $t_3$ (black dotted line) as functions of the rotation angle $\phi$.
}
\label{fivepiecepulse}
\end{figure}

\begin{figure}
\centering
\includegraphics[width=0.7\columnwidth]{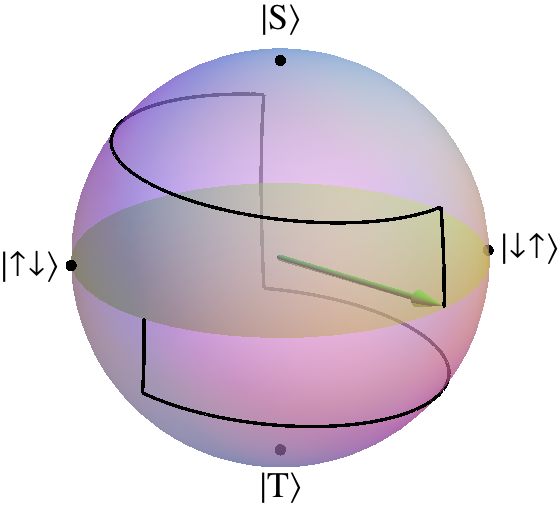}
\caption{Example of evolution of a given state on the Bloch sphere under the five-piece \textsc{supcode}, rotating around $\hat{z}$ by  $\pi/2$. Here $h_0=0$ and for the purpose of illustration we have taken $\delta h/J_{\text{max}} = 0.05$.  $|\!\uparrow\downarrow\rangle$ labels the state with the spin-up 
(spin-down) electron occupying the left (right) dot. Likewise, 
$|\!\downarrow\uparrow\rangle$ labels the spin-permuted state. $|\mathrm{S}\rangle$ denotes the singlet state $\left(|\!\uparrow\downarrow\rangle-|\!\downarrow\uparrow\rangle\right)/\sqrt{2}$, while $|\mathrm{T}\rangle$ denotes the non-magnetic triplet state $\left(|\!\uparrow\downarrow\rangle+|\!\downarrow\uparrow\rangle\right)/\sqrt{2}$. The green arrow is the Bloch vector pointing to the final state after the evolution, which is represented as a point on the surface of the Bloch sphere. The black curve shows the trajectory of the state under the five-piece \textsc{supcode}.}
\label{blochevol5piecePidiv2}
\end{figure}

\section*{Results}
\textbf{Model.}
We consider the Hamiltonian governing the singlet-triplet qubit,
\begin{equation}
H(t)=\frac{h}{2}\sigma_x+\frac{J(t)}{2}\sigma_z,\label{ham}
\end{equation}
with constraints on the parameters imposed to account for the physical realities of the experiments \cite{Petta.05,Foletti.09,Barthel.10,Bluhm.11,Lai.11,Maune.12,Prance.12}.  Here $h=g\mu_B\Delta B$ and fluctuations in $\Delta B$ are much slower than typical gate times so that $h=h_0 + \delta h$ with $h_0$ a known constant and $\delta h$ a random, unknown constant.  ($T^{\ast}_2$ is inversely proportional to the width of the statistical distribution from which $\delta h$ is drawn.  We estimate the effect of high-frequency noise components in the Supplementary Discussion.)  The qubit is manipulated via the electrically controlled exchange coupling, $J(t)$, which is constrained to be positive (except in very high magnetic fields, when it is always negative) and is restricted in magnitude either by the practice of keeping the qubit near the zero-bias point to
reduce charge noise sensitivity, or, more intrinsically, by the
singlet-triplet splitting of two electrons on a single dot,
so that $0\leq J(t) \leq J_{\text{max}}$. The constraint to positive rotations is also relevant to exchange-only coded qubits \cite{DiVincenzo.00,Laird.10,Shi.11}.

Our goal is to design a pulse in $J(t)$ that respects these constraints and performs a given rotation in a way that is insensitive to $\delta h$.  We dub the resulting composite pulse sequence {\sc supcode}: soft uniaxial positive control for orthogonal drift error.  We emphasize that although the singlet-triplet qubit is one of the most experimentally advanced paths towards a scalable quantum computer, the restricted control available does not permit application of existing elegant methods of quantum control.
In particular, the positivity of $J(t)$ precludes the prescriptions of Refs.~\cite{Goelman.89,Grace.11} (which anyway do not accommodate universal single-qubit operations) and   \cite{Khodjasteh.09}.
While other works allow a positivity constraint \cite{Khodjasteh.10,Bensky.10}, they are nonetheless precluded by the uncontrolled, always-on gradient $h$.
Below we consider two cases: $h_0=0$, which is directly relevant to experiments with unpumped nuclear spins and no micromagnet \cite{Petta.05,Maune.12} where only rotations about the $z$-axis are desired (\textit{e.g.}, for spin echo), and $h_0\sim J_{\text{max}}$, which is directly relevant to experiments with pumped nuclear spins \cite{Foletti.09} or a nearby micromagnet \cite{Brunner.11} where full single-qubit control is desired.

\textbf{{\footnotesize{SUPCODE}} for \boldmath $h_0=0$.}
For $h_0=0$, we assume for simplicity that the pulse is of a binary form where $J(t)$ alternates between its extremal values of $0$ and $J_{\text{max}}$.  Expanding the evolution operator of the system, $U\left(T\right) = \mathcal{T}\left\{\exp\left[-i\int_{0}^Tdt H\left(t\right) \right]\right\}$, in powers of $\delta h/J_{\text{max}}$, we choose the time duration of each segment such that the zeroth order term is a rotation about $\hat{z}$ by the desired angle and one or more successively higher order terms vanish.  We give details in the Methods, and in Supplementary Discussion we also rigorously show that there is no pulse which cancels all higher orders for finite $J_{\text{max}}$.  Thus, we find three-, five-, seven-, and nine-piece {\sc supcode} pulses that perform arbitrary rotations about $\hat{z}$ while canceling undesired terms up to the first, second, second, and third order in $\delta h/J_{\text{max}}$, respectively. (We note that in this special case, the first order cancellation could also be performed via Ref.~\cite{Bensky.10}.) An example of the five-piece {\sc supcode} pulse, which cancels first and second order terms in $\delta h/J_{\text{max}}$ in the evolution operator (corresponding to second and fourth order terms in the gate error), is shown in Fig.~\ref{fivepiecepulse}.  Figure \ref{blochevol5piecePidiv2} traces the evolution of a particular initial state on the Bloch sphere under a five-piece {\sc supcode} pulse designed to perform a $\pi/2$ rotation about $\hat{z}$.

\begin{figure}
\centering
\includegraphics[width=0.95\columnwidth]{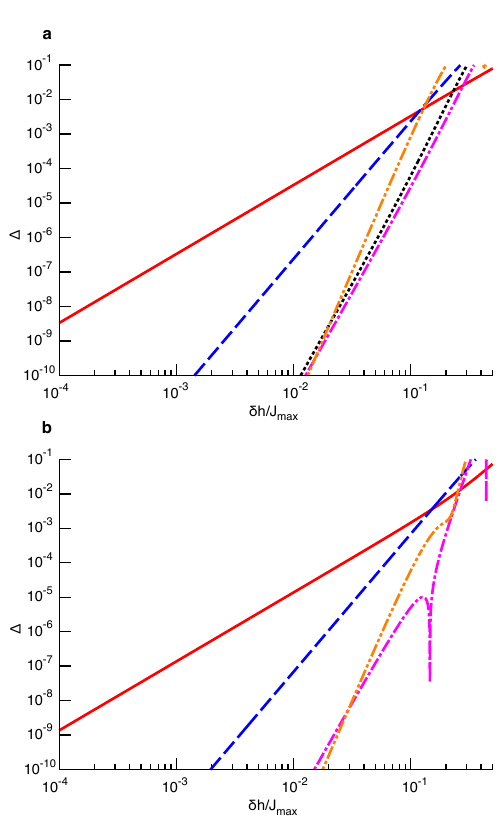}
\caption{Average error per gate for \textsc{supcode} pulses at $h_0=0$. These figures show the average error per gate $\Delta$ as functions of $\delta h/J_{\text{max}}$ for the rotation around the $z$-axis by (\textbf{a}) $\phi_0=\pi/2$ and (\textbf{b}) $\phi_0=1.7\pi$. The red solid curves are for the uncorrected pulse with quadratic leading order errors in $\delta h$. Other curves are showing the three-piece (blue dashed lines), five-piece (black dotted line), seven-piece (magenta dash-dotted lines), and nine-piece (orange dash-dotted lines)    
 \textsc{supcode} pulses
 that cancel gate error $\Delta$ up through second, fourth, fourth, and sixth order in $\delta h$, respectively. For details of the pulses, see Methods and Supplementary Information.}
\label{epghzero}
\end{figure}

\begin{figure}
\centering
\includegraphics[width=0.95\columnwidth]{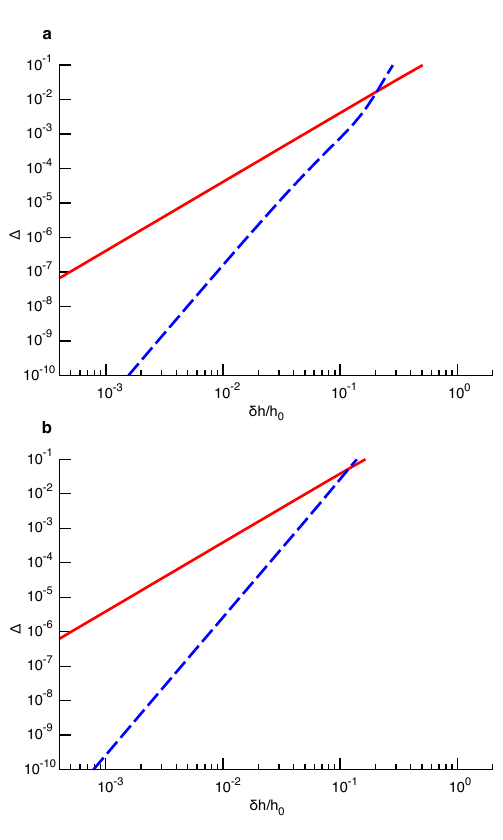}
\caption{Average error per gate for \textsc{supcode} pulse at
  $J_{\text{max}}/h_0\geq 5$.
 This set of figures shows the average error per gate $\Delta$ as functions of $\delta h/h_0$, for rotations by $\pi/2$ about the (\textbf{a})~$x$-axis and (\textbf{b})~$z$-axis. The red curves show the error per gate corresponding to the uncorrected pulses. The corrected pulses, shown by the blue dashed lines, cancel the leading order error. Note that in this figure we used $h_0$ as the energy scale instead of $J_{\text{max}}$.
\label{fig:finiteh}}
\end{figure}

We define the average error per gate, $\Delta$, as
\begin{align}
\Delta=1-\overline{|\langle\psi_i |V^\dagger U(T_f)|\psi_i\rangle|^2}\label{errorpergate}
\end{align}
where $V$ is the desired operation, $U(T_f)$ is the actual evolution operator under the composite pulse of duration $T_f$, and the overlap is averaged over initial states $|\psi_i\rangle$ distributed uniformly over the Bloch sphere. Figure \ref{epghzero} shows the average error per gate for the na\"{i}ve one-piece pulse and the {\sc supcode} pulses introduced above. We see that for $\delta h/J_{\text{max}} < 10\%$ the {\sc supcode} pulses have markedly less error. This range is relevant to recent landmark experiments with GaAs \cite{Bluhm.11} and silicon-based \cite{Maune.12} double quantum dots.  In the former $\delta h \sim 8$neV and in the latter $\delta h \sim 3$neV while for both $J_{\text{max}}$ could be several hundred neV. We also see that although the leading order of the error for the nine-piece pulse is higher than that of the other pulses, its coefficient is large enough that for $\delta h/J_{\text{max}} \gtrsim 2\%$, it is better to use the five- or seven-piece pulses. This suggests that generating similar pulses with even more pieces likely will not extend the range of values of $\delta h$ for which low-error rotations are possible.

\textbf{{\footnotesize{SUPCODE}} for \boldmath $h_0\sim J_{\text{max}}$.}
For $h_0\sim J_{\text{max}}$, the noncommutation of the Hamiltonian at different times even for $\delta h = 0$ makes the previous approach algebraically forbidding. However, in this case we have the possibility to perform rotations about axes $\hat{r}$ other than $\hat{z}$.  We make use of this freedom in order to construct {\sc supcode} pulses in a simple way: We first take an uncorrected (i.e., designed as if $\delta h = 0$) rotation $\tilde{R}(\hat{r};\phi)$ and then construct an uncorrected identity operation $\tilde{I}(\hat{r};\phi)$ designed such that the error in its implementation exactly cancels the leading order error in the original rotation. We represent the uncorrected rotations $\tilde{R}(\hat{r};\phi)$ in terms of ideal rotations $R(\hat{r};\phi)$ as
\begin{equation}
\tilde{R}(\hat{r};\phi)=R(\hat{r};\phi)\bigl(I+\delta h\, A(\hat{r};\phi) + \mathcal{O}(\delta h)^2\bigr)
\end{equation}
and
\begin{equation}
\tilde{I}(\hat{r};\phi)=I+\delta h\, B(\hat{r};\phi) + \mathcal{O}(\delta h)^2,
\end{equation}
where $A(\hat{r};\phi)$ and $B(\hat{r};\phi)$ are the first order corrections. For the corrected pulse
$\tilde{R}\tilde{I}=R\bigl(I+\delta h (A+B) + \mathcal{O}(\delta h)^2\bigr)$, it is clear that we want to make $A(\hat{r};\phi)+B(\hat{r};\phi)=0$. Because there are infinitely many ways to realize the identity operation, there is sufficient freedom to engineer such a cancellation. In particular, we consider parameterized versions of the identity implemented as nested interrupted $2n\pi$ rotations about different axes, with two simple examples being
\begin{equation}
\tilde{I}_1(a_i,b_i)=\tilde{R}(\hat{x}+b_i \hat{z};a_i)\tilde{R}(\hat{x};2\pi)\tilde{R}(\hat{x}+b_i\hat{z};2\pi-a_i)
\end{equation}
and
\begin{equation}
\tilde{I}_2(a_i,b_i,n)=\tilde{R}(\hat{x};a_i)\tilde{R}(\hat{x}+b_i\hat{z};2n\pi)\tilde{R}(\hat{x};2\pi-a_i) ,
\end{equation}
where $a_i$ and $b_i$ are parameters satisfying $0\le a_i\le2\pi$, $0\le b_i\le J_\text{max}/h_0$ and $n$ is a positive integer.

As an explicit demonstration of {\sc supcode}, we show rotations around the $x$- and $z$-axes by arbitrary angles $0<\phi<\pi$ for the particular case $J_{\text{max}}/h_0\geq 5$. (Combinations of these are sufficient for universal one-qubit gates.)  About $\hat{x}$, the uncorrected rotation is performed by holding $J=0$ for a time $\phi/h_0$, and this is preceded by the identity $\tilde{I}=\tilde{I}_1(1/2,b_1)^2\tilde{I}_2(1/2,b_2,15)$ where $b_1$ and $b_2$ are chosen such that errors cancel (see Methods).  About $\hat{z}$, the uncorrected rotation is performed by a three-part pulse \cite{HansonBurkard.07,Ramon.11} $\tilde{R}(\hat{z};\phi)=\tilde{R}(\hat{x}+\hat{z};\pi)\tilde{R}(\hat{x};\phi)\tilde{R}(\hat{x}+\hat{z};\pi)$, and this is preceded by the identity $\tilde{I}=\tilde{I_1}(1/2,b_3)^3\tilde{I_2}(a,b_4,15)$ where $a$, $b_3$, and $b_4$ are chosen such that errors cancel (see Methods). As shown in Fig.~\ref{fig:finiteh} for $\phi=\pi/2$, {\sc supcode} does indeed lead to a higher-order scaling of the error in $\delta h/h_0$ (and hence in $\hbar/h_0T^{\ast}_2$ ), and a reduction of error when $\delta h/h_0$ is less than a few percent.  For instance, when $\delta h/h_0 = 5\%$, errors are typically reduced by an order of magnitude.

\section*{Discussion}

The tradeoff of our approach is that {\sc supcode} rotations are typically over an order of magnitude longer than uncorrected rotations.  We note that for a given experimental set of parameters one should try optimizing the pulse sequence, which we have not done for the arbitrarily-chosen example above, as both the length and error of the corrected pulses could certainly be reduced by a significant constant factor by searching over different constructions of the identity, $\tilde{I}$. Nonetheless, for experiments in GaAs systems with pumped nuclear spins \cite{Bluhm.10} ($h_0 \sim 0.6\mu$eV, $\delta h \sim 30$neV) or micromagnets \cite{Brunner.11} ($h_0 \sim 20$neV, $\delta h \sim 1$neV) the sequences shown here could already deliver substantial improvement over the uncorrected ones.

Realistic deviations from the ideal pulses assumed above would include charge noise, which adds a random quasi-static contribution to the exchange, and finite rise times.  The former could be reduced by using a multi-electron variant of the singlet-triplet qubit \cite{Barnes.11} and, in principle, it may even be possible to dynamically correct by adding more degrees of freedom to the pulses.  The latter can be compensated for by adjusting pulse parameters given the actual turn-on/off profiles of the pulse for a specific experimental setup (we give explicit demonstrations in the Supplementary Discussion). Thus, even in nonideal conditions {\sc supcode} could enable precise spin qubit rotations independent of shot-to-shot variation in the nuclear Overhauser field, also easing tasks such as ensemble-averaged measurements of singlet probability oscillations versus time by reducing hyperfine-induced decay.  More importantly, this work allows satisfaction of the quantum error correction threshold within a substantially larger region of the physical parameter space than would otherwise be possible.
The fact that $T_2$ has now reached tens of microseconds in GaAs quantum dots \cite{Bluhm.11}, and milliseconds \cite{Witzel.10} or even seconds in the presence of isotopic purification \cite{Tyryshkin.11} in Si-based structures implies that gate errors are currently dominated not by dynamical fluctuations in the nuclear spin bath, but by the statistical distribution of the magnetic field gradient, and these errors may be efficiently suppressed by {\sc supcode}.

\onecolumngrid

\section*{Methods}

\textbf{Expansion of the evolution operator around \boldmath $h_0=0$.}
The evolution operator is defined as
\begin{align}
U\left(T,0\right) = \mathcal{T}\left\{\exp\left[-i\int_{0}^Tdt\left(\frac{h}{2}\sigma_x+\frac{J(t)}{2}\sigma_z\right) \right]\right\},\label{Uoperator}
\end{align}
with $h=h_0+\delta h$.
Expanding the evolution operator in powers of $\delta h$ in the vicinity of $h_0=0$,
\begin{align}
U\left(T, 0\right) = \sum_{n=0}^\infty \delta h^n\Pi_n,
\end{align}
where one can show that
\begin{equation}
\Pi_0 = \cos\left[f(T)\right] I - i \sin\left[f(T)\right]\sigma_z
\end{equation}
and, for $n>0$,
\begin{equation}
\begin{split}
\Pi_n = \left(-\frac{i}{2}\right)^n \left(\prod_{m=n}^1
\int_{0}^{t_{m+1}'}dt_m'\right) \biggl\{\cos\left[\frac{f\left(\tau\right)}{2}-
\sum_{k=1}^n \left(-1\right)^{k-n} f\left(t_k'\right) \right]A_n\\
+
\sin\left[\frac{f\left(\tau\right)}{2}-\sum_{k=1}^n\left(-1\right)^{k-n}
f\left(t_k'\right)\right]B_n \biggr\},\label{nthordercorrection}
\end{split}
\end{equation}
(note that the product is in descending order in observance of the time-ordering of operators)
where
\begin{equation}
t_{n+1}' \equiv T,
\end{equation}
\begin{align}
f(T)=\int_0^{T}dtJ(t),
\end{align}
\begin{equation}
A_n \equiv \begin{cases} I & n\text{ even} \\ \sigma_x & n\text{ odd}
\end{cases}, \qquad B_n \equiv \begin{cases} -i \sigma_z & n\text{ even}
\\ \sigma_y & n\text{ odd} \end{cases},
\end{equation}
and $I$ is the $2\times2$ identity matrix and $\sigma_x$, $\sigma_y$, and $\sigma_z$ are Pauli matrices.

Thus, for the $n$th-order term to vanish, one must have
\begin{equation}
\left(\prod_{m=n}^1 \int_{0}^{t_{m+1}'}dt_m'\right)
\exp\left[i\sum_{k=1}^n \left(-1\right)^kf\left(t_k'\right)\right] = 0.\label{eq:nthcond}
\end{equation}
The no-go theorem shown in Supplementary Discussion implies that it is instructive to design pulses which cancel successive orders. In this work we focus on the piecewise constant pulse, which can be expressed as
\begin{align}
&U\left(T_f,0\right)=\prod_{k=N}^1\exp\left[-i\left(\frac{\delta h}{2}\sigma_x+\frac{J_k}{2}\sigma_z\right)\frac{\tau_k}{J_{\rm max}}\right],\label{Udef:tau}
\end{align}
where we have defined the dimensionless quantity $\tau_k=J_{\rm max} t_k$ for convenience.

Let us also define some simplifying notations. $t_k$ refers to the duration of a pulse on the $k^{\rm th}$ piece. The total duration of time after the $k^{\rm th}$ piece would be $T_k=\sum_{j=1}^kt_j$, with $T_0\equiv 0$ and $T_N\equiv T_f$ indicating the initial and final time. $J(t)$ can then be expressed as
\begin{align}
J(t)=\sum_{k=1}^NJ_k\Theta(t-T_{k-1})\Theta(T_{k}-t),
\end{align}
where $\Theta(t)$ is the Heaviside step function.

\vspace{0.2cm}
\textbf{Three-piece {\footnotesize{SUPCODE}} for \boldmath $h_0=0$.}
Our motivation is to cancel successive orders of $\delta h$ in the expansion with number of pieces $N$ as small as possible. We consider symmetric pulses, i.e., $J(t)=J(T_f-t)$, for the $h_0=0$ case. This ensures that $U(T_f,0)$ has no $\sigma_y$ component. [To see this fact, note that the operator $\exp[-i(h\sigma_x/2+J_k\sigma_z/2)t_k]$ can be written in the form
$A_i=a_{i0}{I}+a_{ix}{\sigma_x}+a_{iz}{\sigma_z}$
with $a_{i0}$, $a_{ix}$, $a_{iz}$ arbitrary complex numbers. It is then straightforward to show that for any operators $A_1$ and $A_2$ with arbitrary coefficients, $A_2\cdot A_1\cdot A_2$ can also be written in such a form, free of $\sigma_y$ terms.
Applying this statement recursively to Eq.~\eqref{Udef:tau}, one sees that for any $J(t)$ satisfying $J(t)=J(T_f-t)$ the resulting evolution operator $U$ does not contain $\sigma_y$ component.]

One of the simplest ways to cancel the leading order error with strictly positive values of $J$ is via a three-piece pulse sequence. In Eq.~\eqref{Udef:tau} we take $N=3$ and the three-piece pulse sequence can be characterized by
\begin{equation}
(\tau_1, J_1)=(\tau_3, J_3)=(\phi_0, J_{\rm max}); \; (\tau_2, J_2)=(4\pi-2\phi_0, J_{\rm max}/2),
\end{equation}
where $\phi_0$ is the desired rotation angle around the $z$-axis. It is straightforward to verify that the evolution operator under this pulse is
\begin{align}
\begin{split}
U(T_f,0)=&-\exp\left(-i\frac{\phi_0}{2}{\sigma_z}\right)\\
&+\frac{\delta h^2}{2J_{\rm max}^2}\left(4\pi-\phi_0+\sin\phi_0\right)\left(\sin\frac{\phi_0}{2}{I}+i\cos\frac{\phi_0}{2}{\sigma_z}\right)+{\cal O}\Big[\Big(\frac{\delta h}{J_{\rm max}}\Big)^3\Big].
\end{split}
\end{align}
Here the deviation from the desired rotation has been suppressed up to second order in $\delta h$ (which corresponds to the fourth order in Eq.~\eqref{errorpergate}). Note that this pulse actually sweeps the Bloch vector through an angle $\phi=2\pi+\phi_0$, which is the origin of the trivial additional phase factor. It is clear that in order to achieve error cancellation, the Bloch vector generally must be swept through more than $2\pi$ about the $z$-axis since its path is deflected from the ideal ($\delta h=0$) path in opposite directions in the ``eastern and western hemispheres." (One can also explicitly show the necessity of larger angles from Eq.~\eqref{nthordercorrection}.) Thus the three-piece pulse sequence cancels leading order error simply by ensuring that, in the absence of $\delta h$, the Bloch vector spends an equal amount of time in each hemisphere during its rotation.

\vspace{0.2cm}
\textbf{Five-piece {\footnotesize{SUPCODE}} for \boldmath $h_0=0$.} We set $N=5$ in Eq.~\eqref{Udef:tau}, which allows us cancel the dependence of $U(T_f,0)$ on $\delta h$ up to the second order (corresponding to the fourth order in Eq.~\eqref{errorpergate}), using a symmetric pulse with parameters $J_1=J_3=J_5=0$, $J_2=J_4=J_{\rm max}$, and
\begin{align}
\tau_1=\tau_5,\quad \tau_2=\tau_4=\phi/4.\label{5pieceothertaus}
\end{align}
We expand $U(T_f,0)$ as
\begin{align}
U&(T_f,0)=\Big(\cos\frac{\phi}{2}{I}-i\sin\frac{\phi}{2}{\sigma_z}\Big)-i\frac{\delta h}{2J_{\rm max}}\Big(\tau_3+2\tau_1\cos\frac{\phi}{2}+2\sin\frac{\phi}{2}\Big){\sigma_x}\notag\\
&-\frac{\delta h^2}{8J_{\rm max}^2}\Bigg\{\Big[4\tau_1\tau_3+(4\tau_1^2+\tau_3^2)\cos\frac{\phi}{2}+2(4\tau_1+2\tau_3+\phi)\sin\frac{\phi}{2}\Big]{I}\notag\\
&\qquad\qquad\ -i\Big[4\tau_3-2(2\tau_3+\phi)\cos\frac{\phi}{2}+(4+\tau_3^2)\sin\frac{\phi}{2}\Big]{\sigma_z}\Bigg\}\notag\\
&+{\cal O}\Big[\Big(\frac{\delta h}{J_{\rm max}}\Big)^3\Big].\label{5pieceU}
\end{align}

To make the first order coefficient vanish, one must choose
\begin{align}
\tau_3=-2\left(\tau_1\cos\frac{\phi}{2}+\sin\frac{\phi}{2}\right).\label{5piecetau3}
\end{align}

Plugging Eq.~\eqref{5piecetau3} into Eq.~\eqref{5pieceU}, it suffices to satisfy
\begin{align}
-2\tau_1-\phi+4\tau_1\cos\frac{\phi}{2}-2\tau_1\cos\phi+4\sin\frac{\phi}{2}+(\tau_1^2-1)\sin\phi=0
\end{align}
to make the second order terms vanish.
We choose a root that is positive for $2\pi<\phi<3\pi$, that is
\begin{align}
\tau_1&=\csc\phi\left(1-2\cos\frac{\phi}{2}+\cos\phi+\sqrt{4-8\cos\frac{\phi}{2}+4\cos\phi+\phi\sin\phi}\right).\label{5piecetau1}
\end{align}

Equations~\eqref{5pieceothertaus}, \eqref{5piecetau3}, and \eqref{5piecetau1} prescribe the pulse parameters required to achieve a $R(\hat{z};\phi)$ rotation while canceling the dependence of $U(T_f,0)$ on $\delta h$ up to the second order.
A plot of these parameters as a function of $\phi$ is given in Fig~\ref{fivepiecepulse}. Note that this pulse is defined for $\phi\in(2\pi,3\pi)$, which is equivalent to a rotation $R(\hat{z};\phi_0)$ with $\phi_0\in(0,\pi)$. Rotation of angles outside the range $(0,\pi)$ may be achieved by duplicating existing pulse sequences. For example, to achieve a $\hat{z}$-rotation of $\phi_0=1.2\pi$ one could apply twice the $\phi_0=0.6\pi$ (corresponding to $\phi=2.6\pi$) rotation.

As $\phi\rightarrow2\pi$, $\tau_1\rightarrow\infty$, $\tau_2=\pi/2$, $\tau_3\rightarrow2\tau_1$. If we want to fix the total duration of the sequence, we let $J_{\rm max}\rightarrow\infty$. The pulse sequence becomes the well-known CPMG pulse \cite{Carr.54,Meiboom.58}. In fact, it can be shown that for a pair of instantaneous $\pi$ pulses, setting leading order errors to zero while maintaining the time-reversal symmetry of the pulse sequence enforces the Uhrig condition \cite{Uhrig.07}.

For details of the construction of seven-piece and nine-piece \textsc{supcode}, see Supplementary Methods.

\vspace{0.2cm}
\textbf{{\footnotesize SUPCODE} for \boldmath $h_0\neq0$.}
As described in Results, we perform corrected rotations as $\tilde{R}\tilde{I}$ where the implementation of the identity $\tilde{I}$ is chosen such that the first order errors in $\delta h$ exactly cancel those from the rotation $\tilde{R}$. Our implementations of $\tilde{I}$ are of the form
\begin{equation}
    \tilde{I}(\hat{r};\phi)=\prod_i\tilde{I}_{n_i}(a_i,b_i) \label{iform},
\end{equation}
For the specific forms $\tilde{I}_1$, $\tilde{I}_2$ given in the main text, the parameters $a_i$, $b_i$ are constrained by
\begin{align}
    &0\leq a_i\leq 2\pi ,\label{constraint0}\\
    &0\leq b_i\leq J_{\text{max}}/h_0 \label{constraint} .
\end{align}
The particular implementation of $\tilde{I}$ that will be chosen for a given experimental situation will depend on the particular rotation $\tilde{R}$ as well as the experimental parameters $J_{\text{max}}/h_0$, with a combinatorial search necessary to find an implementation that: (a)~cancels the first-order error from $\tilde{R}$; (b)~satisfies the constraints of Eqs.~\eqref{constraint0} and \eqref{constraint}; and (c)~is close to `optimal' in some experimentally meaningful sense, such as having the overall shortest duration in time, having the smallest second-order error in $\delta h$, or being least sensitive to over-/under-shoot in $J(t)$.  As an example of our technique we found sequences valid for the experimentally relevant regime $J_{\text{max}}/h_0 \geq 5$, for corrected rotations about angles $0\leq\phi\leq\pi$ around the $x$- and $z$-axes. (We have also verified that slightly more complicated identities can be constructed to correct error for pulses with smaller values of $J_{\text{max}}/h_0>1.5$.) For the $x$-axis rotation we have
\begin{equation}
    R(\hat{x};\phi)+ \mathcal{O}(\delta h)^2 = \tilde{R}(\hat{x};\phi)\tilde{I}_1(1/2,b_1)^2\tilde{I}_2(1/2,b_2,15) .
\end{equation}
The first-order term in $\delta h$ is zero when $b_{1,2}$ are chosen to satisfy
\begin{gather}
\frac{\phi }{2 \pi }=-\frac{2 h_0^3}{\bigl(h_0^2+b_1^2\bigr)^{3/2}}-
\frac{15 h_0^3}{\bigl(h_0^2+b_2^2\bigr)^{3/2}}+
\frac{4 b_1^2}{h_0^2+b_1^2}-3 ,\\
\frac{2 \sqrt{h_0^2+b_1^2}+h_0}{\bigl(h_0^2+b_1^2\bigr)^{3/2}}=
\frac{15 b_2 h_0}{2 b_1 \bigl(h_0^2+b_2^2\bigr){}^{3/2}} ,
\end{gather}
 The solution to these equations are shown in Supplementary Figure S3. In Supplementary Figure S3\textbf{a} we show a specific example of the pulse sequence for $\pi/2$ rotation about the $x$-axis. Supplementary Figure S3\textbf{c} shows solutions for a range of rotation angle $0\le\phi\le\pi$. 

For the same parameter range we construct the corrected $z$-axis rotation as
\begin{equation}
   R(\hat{z};\phi)+\mathcal{O}(\delta h)^2= \tilde{R}(\hat{x}+\hat{z};\pi)\tilde{R}(\hat{x};\phi)\tilde{R}(\hat{x}+\hat{z};\pi)\tilde{I_1}(1/2,b_3)^3\tilde{I_2}(a,b_4,15),
\end{equation}
with $a$ and $b_{3,4}$ satisfying
\begin{gather}
\frac{6 h_0^3}{\bigl(h_0^2+b_3^2\bigl)^{3/2}}+
\frac{30 h_0^3}{\bigr(h_0^2+b_4^2\bigr)^{3/2}}+
\frac{12 h_0^2}{h_0^2+b_3^2}+
\frac{\sin\phi}{\pi }+
\frac{\cos\phi+1}{2 \sqrt{2}}=4 ,\\
\cos\phi=\frac{30 \pi  b_4 h_0^2 \sin a}{\bigl(h_0^2+b_4^2\bigr)^{3/2}}+
\frac{\pi  \sin\phi}{2 \sqrt{2}}+1 ,\\
\frac{30 b_4 h_0 \cos a}{\bigl(h_0^2+b_4^2\bigr)^{3/2}}+
\frac{12 b_3}{h_0^2+b_3^2}+
\frac{6 b_3 h_0}{\bigl(h_0^2+b_3^2\bigr)^{3/2}}+
\frac{\phi }{\pi  h_0}+\frac{1}{\sqrt{2} h_0}=0 .
\end{gather}
The solution to these equations is shown in Supplementary Figure S4. In Supplementary Figure S4\textbf{a} we show an example of the pulse sequence appropriate for rotating about the $z$-axis by $\pi/2$, while in Supplementary Figure S4\textbf{c} we show solutions for $0\le\phi\le\pi$.

\vspace{0.5cm}

This work is supported by LPS-NSA-CMTC and IARPA-MQCO grants. The authors declare that they have no
competing financial interests. Correspondence and requests for materials
should be addressed to Xin Wang (email: xin@umd.edu).

\twocolumngrid




\onecolumngrid

\vspace{0.5cm}

\setcounter{secnumdepth}{3}  
\setcounter{equation}{0}
\setcounter{figure}{0}
\renewcommand{\theequation}{S-\arabic{equation}}
\renewcommand{\thefigure}{S\arabic{figure}}
\renewcommand\figurename{Supplementary Figure}
\newcommand\Scite[1]{[S\citealp{#1}]}

\makeatletter \renewcommand\@biblabel[1]{[S#1]} \makeatother


\section*{Supplementary methods}

\subsection{Seven-piece {\footnotesize SUPCODE} for \boldmath $h_0=0$}\label{sevenpiece}

In Eq.~\eqref{Udef:tau} we take $N=7$,  $J_1=J_3=J_5=J_7=0$, $J_2=J_4=J_6=J_{\rm max}$, and
\begin{equation}
\tau_1=\tau_7,\quad \tau_2=\tau_4=\tau_6=\phi/3,\quad \tau_3=\tau_5.\label{sevenpieceothertau}
\end{equation}
Since the number of independent variables are the same as for the five-piece pulse, it can still cancel the dependence of $U(T_f,0)$ on $\delta h$ up to second order, but not to third order. However, it will expand the range of rotation angles for which the pulse can directly be applied. We expand $U(T_f,0)$ as
\begin{align}
&U(T_f,0)=\Big(\cos\frac{\phi}{2}{I}-i\sin\frac{\phi}{2}{\sigma_z}\Big)-i\frac{\delta h}{J_{\rm max}}\Big(\tau_3\cos\frac{\phi}{6}+\tau_1\cos\frac{\phi}{2}+\sin\frac{\phi}{2}\Big){\sigma_x}\notag\\
&-\frac{\delta h^2}{4J_{\rm max}^2}\Bigg\{\Big[\tau_3(4\tau_1+\tau_3)\cos\frac{\phi}{6}+(2\tau_1^2+\tau_3^2)\cos\frac{\phi}{2}+(\phi+4\tau_1+4\tau_3)\sin\frac{\phi}{2}\Big]{I}\notag\\
&\qquad\qquad-i\Bigg[4\tau_3\cos\frac{\phi}{6}-(4\tau_3+\phi)\cos\frac{\phi}{2}+\tau_3^2\sin\frac{\phi}{6}+(\tau_3^2+2)\sin\frac{\phi}{2}\Bigg]{\sigma_z}\Bigg\}\notag\\
&+{\cal O}\Big[\Big(\frac{\delta h}{J_{\rm max}}\Big)^3\Big].\label{7pieceU}
\end{align}

Similar to the previous section, to make the first order coefficient vanish, one must choose
\begin{align}
\tau_3=-\sec\frac{\phi}{6}\Big(\tau_1\cos\frac{\phi}{2}+\sin\frac{\phi}{2}\Big).\label{7piecetau3}
\end{align}

Plugging Eq.~\eqref{7piecetau3} into Eq.~\eqref{7pieceU}, it suffices to satisfy
\begin{align}
\begin{split}
8\cos\frac{\phi}{6}-6\cos\frac{\phi}{2}-&(\tau_1^2-5)\cos\frac{5\phi}{6}
+(\tau_1^2-1)\cos\frac{7\phi}{6}-6\sec\frac{\phi}{6}\\
+&\Bigg[
8\tau_1+\phi+4\tau_1\Big(\cos\frac{2\phi}{3}-3\cos\frac{\phi}{3}\Big)
\Bigg]\sin\frac{\phi}{2}
=0
\end{split}
\end{align}
to make the second order terms vanish.
We choose a root that is positive for $3\pi<\phi<5\pi$, that is
\begin{align}
\tau_1=\frac{2\cos\frac{\phi}{2}-\cos\frac{\phi}{6}-\cos\frac{5\phi}{6}+2\sqrt{\cos^2\frac{\phi}{6}
\left(\phi\cos\frac{\phi}{6}-6\sin\frac{\phi}{6}\right)
\left(\sin\frac{\phi}{2}-2\sin\frac{\phi}{6}\right)
}}{\sin\frac{\phi}{6}-\sin\frac{5\phi}{6}}.\label{7piecetau1}
\end{align}

Equations~\eqref{sevenpieceothertau}, \eqref{7piecetau3}, and \eqref{7piecetau1} give the pulse parameters of seven-piece \textsc{supcode} capable of achieving a $R(\hat{z};\phi)$ rotation while canceling the dependence of $U(T_f,0)$ on $\delta h$ up to second order. A plot of these parameters as a function of $\phi$ is given in Supplementary Figure~\ref{para7piece}. Note that this pulse is defined for $\phi\in(3\pi,5\pi)$, which is equivalent to a rotation $R(\hat{z};\phi_0)$ with $\phi_0\in(-\pi,\pi)$.

\begin{figure}[t]
\centering
\includegraphics[width=9cm]{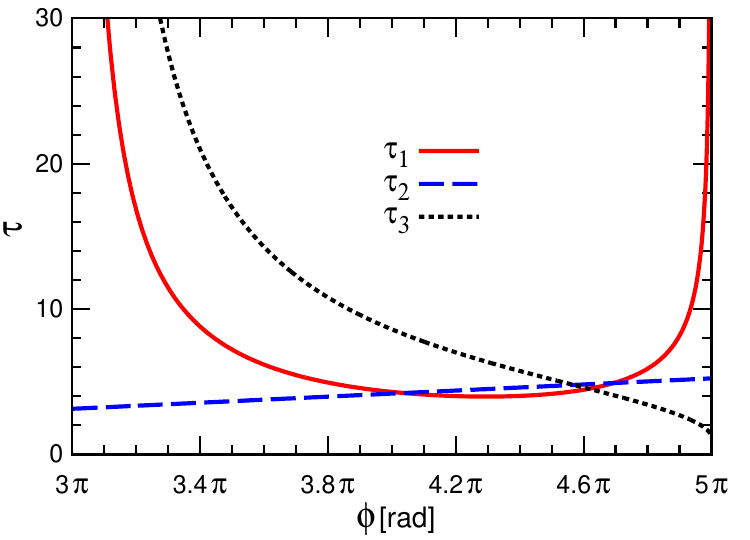}
\caption{{\bf Parameters for the seven-piece {\footnotesize SUPCODE} pulse.} $\tau_1$ (red solid line), $\tau_2$ (blue dashed line) and $\tau_3$ (black dotted line) are parameters for the seven-piece \textsc{supcode} pulse as defined in the Supplementary Methods. This pulse cancels the dependence of the evolution operator on $\delta h$ up to second order, corresponding to the fourth order in the average error per gate.}
\label{para7piece}
\end{figure}

As $\phi\rightarrow3\pi$,  both $\tau_1,\tau_3\rightarrow\infty$, $\tau_2=\tau_4=\pi$. However, in this limit, we have
\begin{equation}
\frac{\tau_3}{\tau_1}\rightarrow\frac{4+3\sqrt{2}}{2+\sqrt{2}}=1+\sqrt{2}.
\end{equation}
Note that $\sin\frac{\pi}{8}=\sqrt{2-\sqrt{2}}/2$, and
\begin{equation}
\frac{\sin^2\frac{2\pi}{8}-\sin^2\frac{\pi}{8}}{\sin^2\frac{\pi}{8}}=1+\sqrt{2}.
\end{equation}
Therefore in this limit the pulse is nothing but the $n=3$ Uhrig pulse, which can be understood in a similar way as in the previous section.

\subsection{Nine-piece {\footnotesize SUPCODE} for \boldmath $h_0=0$}\label{ninepiece}

In Eq.~\eqref{Udef:tau} we take $N=9$ and the nine-piece pulse sequence can be expressed as
$\tau_1=\tau_9$, $\tau_2=\tau_4=\tau_6=\tau_8=\phi/4$, $\tau_3=\tau_7$, $J_1=J_3=J_5=J_7=J_9=0$, $J_2=J_4=J_6=J_8=J_{\rm max}$. Now there is one more independent variable, which means that we are able to cancel the dependence of $U(T_f,0)$ on $\delta h$ up to the third order. The expansion of the evolution operator is complicated, so we start with the first order in $\delta h$, which is
\begin{align}
-i\frac{\delta h}{2J_{\rm max}}\Big(\tau_5+2\tau_3\cos\frac{\phi}{4}+2\tau_1\cos\frac{\phi}{2}+2\sin\frac{\phi}{2}\Big){\sigma_x}.
\end{align}
Making this term vanish requires
\begin{align}
\tau_5=-2\left(\tau_3\cos\frac{\phi}{4}+\tau_1\cos\frac{\phi}{2}+\sin\frac{\phi}{2}\right).\label{9piecetau5}
\end{align}

Using Eq.~\eqref{9piecetau5}, the expansion simplifies to a function only dependent on $\tau_1$, $\tau_3$ and $\phi$. To further simplify the notations we introduce functions
\begin{align}
\begin{split}
f_1(\tau_1,\tau_3,\phi)&=-2\tau_1-4\tau_3-\phi+6\tau_3\cos\frac{\phi}{4}+4\tau_1\cos\frac{\phi}{2}-2\tau_3\cos\frac{3\phi}{4}-2\tau_1\cos\phi\\
&\quad-2\tau_1\tau_3\sin\frac{\phi}{4}+(\tau_3^2+4)\sin\frac{\phi}{2}+2\tau_1\tau_3\sin\frac{3\phi}{4}+(\tau_1^2-1)\sin\phi,\\
\end{split}
\end{align}
and
\begin{align}
f_2(\tau_1,\tau_3,\phi)&=2\tau_3^2\cos\frac{\phi}{4}+2\Big[\tau_1^3+3\tau_1(3+\tau_3^2)+3(4\tau_3+\phi)\Big]\cos\frac{\phi}{2}
-6\tau_1(\tau_3^2+4)\cos\phi\notag\\
&\quad-2\tau_3(15-3\tau_1^2+\tau_3^2)\cos\frac{3\phi}{4}-6(\tau_1^2-1)\tau_3\cos\frac{5\phi}{4}
-2\tau_1(\tau_1^2-3)\cos\frac{3\phi}{2}\notag\\
&\quad-3\tau_3(4\tau_3+\phi)\sin\frac{\phi}{4}-6\Big[-1+\tau_1^2-3\tau_3^2+\tau_1(4\tau_3+\phi)\Big]\sin\frac{\phi}{2}
+36\tau_1\tau_3\sin\frac{3\phi}{4}\notag\\
&\quad+6(2\tau_1^2-\tau_3^2-2)\sin\phi-12\tau_1\tau_3\sin\frac{5\phi}{4}+2(1-3\tau_1^2)\sin\frac{3\phi}{2}.
\end{align}

With these definitions, we express the expansion of the evolution operator as
\begin{align}
&U(T_f,0)=\Big(\cos\frac{\phi}{2}{I}-i\sin\frac{\phi}{2}{\sigma_z}\Big)\notag\\
&+\frac{\delta h^2}{4J_{\rm max}^2}f_1(\tau_1,\tau_3,\phi)\Big[\sin\frac{\phi}{2}{I}+i\cos\frac{\phi}{2}{\sigma_z}\Big]-i\frac{\delta h^3}{24J_{\rm max}^3}f_2(\tau_1,\tau_3,\phi){\sigma_x}\notag\\
&+{\cal O}\Big[\Big(\frac{\delta h}{J_{\rm max}}\Big)^4\Big].\label{9pieceU}
\end{align}

To cancel the dependence of $U(T_f,0)$ on $\delta h$ up to third order, we must find $\tau_1$ and $\tau_3$ as positive, real solutions to the coupled nonlinear equations
\begin{align}
\left\{
\begin{matrix}
f_1(\tau_1,\tau_3,\phi)=0\\
f_2(\tau_1,\tau_3,\phi)=0
\end{matrix}
\right.,
\end{align}
while at the same time requiring a positive $\tau_5$. We found a numerical solution to the above equations for $\phi\in(4\pi,4.5654\pi)\cap(5\pi,6\pi)$, which corresponds to rotation $R(\hat{z};\phi_0)$ with $\phi_0\in(0,0.5654\pi)\cap(\pi,2\pi)$. The numerical solution is shown in Supplementary Figure~\ref{para9piece}.

\begin{figure}[!]
\centering
\includegraphics[width=9cm]{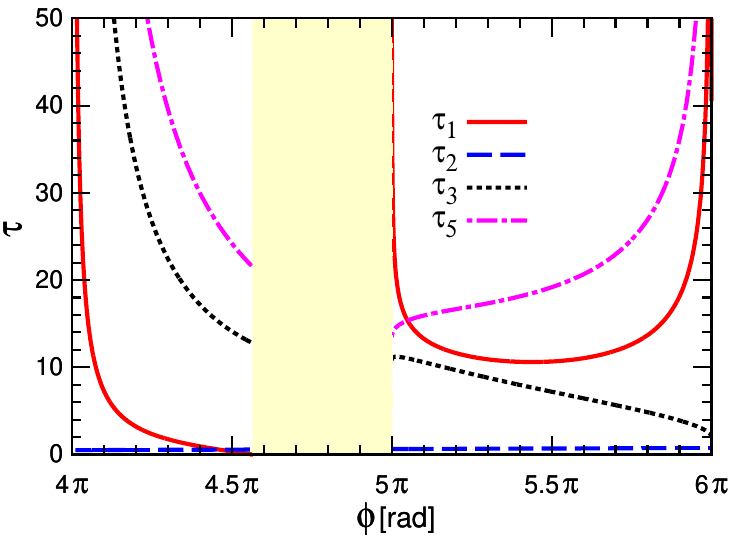}
\caption{{\bf Parameters for the nine-piece {\footnotesize SUPCODE} pulse.} $\tau_1$ (red solid line), $\tau_2$ (blue dashed line), $\tau_3$ (black dotted line) and $\tau_5$ (magenta dash-dotted line) are parameters for the nine-piece \textsc{supcode} pulse as defined in the Supplementary Methods. The area shaded by light yellow indicate the regime where positive $\tau_1$ cannot be found. This pulse cancels the dependence of the evolution operator on $\delta h$ up to third order, corresponding to the sixth order in the average error per gate.}
\label{para9piece}
\end{figure}

\section*{Supplementary discussion}

\subsection{Estimated effect of high-frequency noise}\label{noise}
In reality, the noise is not completely static, but has a spectral density with a high-frequency tail.  A typical model for this tail is
\begin{equation}
S(\omega) = \frac{A^{\beta+1}}{\omega^{\beta}},
\end{equation}
where fitting recent data yields $\beta = 2.6$ and $A^{-1} = 3.6 \mu$s \Scite{Medford12}.  For our pulse sequences of length $T$, the effect of noise with frequency lower than $2\pi/T$ is strongly suppressed by construction.  The effect of the higher frequency noise on the fidelity of a composite rotation depends on the particular pulse sequence.  However, we can roughly estimate the order of magnitude of the effect by considering the free induction decay after time $T$ due to the spectral density of noise above $2\pi/T$.

Following Ref.~\Scite{Cywinski08}, the coherence function, $W(t)=e^{-\chi(t)}$, is determined by
\begin{equation}
\chi(t)=\int_{\omega_c}^{\infty} \frac{d\omega}{\pi} S(\omega) \frac{F(\omega t)}{\omega^2},
\end{equation}
with $F(\omega t)=2 \sin^2(\omega t/2)$.  Switching over to a dimensionless integration variable $\nu=\omega t/2$ and using the explicit form of $S(\omega)$ with $\omega_c = 2\pi/T$, we have
\begin{equation}
\chi(t)=(A t)^{1+\beta} g(\pi t/T),
\end{equation}
where
\begin{equation}
g(x) = 1/(2^\beta\pi)\int_x^{\infty} d\nu \nu^{-2-\beta}\sin^2(\nu).
\end{equation}

If we want the gate error from the high-frequency noise to be less than a threshold value of $10^{-4}$, then $A$, $\beta$, and $T$ must be such that $\chi(T) < 10^{-4}$.  This translates to
\begin{equation}
A T \left(10^4 g(\pi)\right)^{1/(1+\beta)} < 1.
\end{equation}
Assuming $\beta=2.6$, the numerical factor is of order unity and one simply has $A T < 1$.  Further assuming the experimental value of $A^{-1} \sim 3.6 \mu$s \Scite{Medford12}, the condition becomes $T < 3.6\mu$s.  The pulse sequences shown in Figs.~\ref{paraRx} and \ref{paraRz}, even without further optimization for length, satisfy this condition for $h_0 > 15$neV.

\subsection{No-go theorem for canceling error to all orders}\label{nogo}

Here we show that there does not exist a pulse sequence with a bounded exchange interaction that suppresses all orders of error due to $\delta h$.

Consider a function
\begin{align}
\psi(T,\lambda)&=1+\sum_{n=1}^\infty\lambda^n\left(\prod_{m=n}^1 \int_{0}^{t_{m+1}'}dt_m'\right)
\exp\left[i\sum_{k=1}^n \left(-1\right)^kf\left(t_k'\right)\right]\notag\\
&=1+\lambda\int_0^Tdt_1'\exp\left[-if(t_1')\right]+\lambda^2\int_0^Tdt_2'\int_0^{t_2'}dt_1'\exp\left[-if(t_2')+if(t_1')\right]\notag\\
&\quad+\lambda^3\int_0^Tdt_3'\int_0^{t_3'}dt_2'\int_0^{t_2'}dt_1'\exp\left[-if(t_3')+if(t_2')-if(t_1')\right]+\cdots
\end{align}
which solves the partial differential equation
\begin{align}
\frac{\partial \psi}{\partial T}=\lambda\exp\left[-if(T)\right]\psi^*,\label{psieq}
\end{align}
with boundary condition
\begin{align}
\psi(T=0,\lambda)=1.
\end{align}
Observe that if we can find a pulse which cancels all orders in Eq.~\eqref{eq:nthcond} at $T=T_f$, it means that Eq.~\eqref{psieq} has a solution that also satisfies $\psi(T=T_f,\lambda)=1$ for any $\lambda$. To simplify the problem further, we write
\begin{align}
\psi(T,\lambda)=\exp\left[-i\frac{f(T)}{2}\right]\exp\left[u(T,\lambda)-i\frac{w(T,\lambda)}{2}\right],\label{asdfads}
\end{align}
where $u(T,\lambda)$ and $w(T,\lambda)$ are real functions satisfying
\begin{align}
\frac{\partial u(T,\lambda)}{\partial T}&=\lambda\cos w(T,\lambda),\\
\frac{\partial w(T,\lambda)}{\partial T}&=-2\lambda\sin w(T,\lambda)-J(T).\label{eqw}
\end{align}
The associated boundary conditions are
\begin{align}
u(T=0,\lambda)&=0,\\
u(T=T_f,\lambda)&=\lambda\int_0^{T_f}dT'\cos w(T',\lambda)=0,\label{bc:zeroth}\\
w(T=0,\lambda)&=0,\label{bc:first}\\
w(T=T_f,\lambda)&=-f(T_f).\label{bc:second}
\end{align}
We focus on Eqs.~\eqref{eqw},~\eqref{bc:zeroth}-\eqref{bc:second} only and consider the limit $\lambda\rightarrow\infty$. It is clear from Eq.~\eqref{eqw} that $w(T,\lambda)$ exists in this limit and is given by
\begin{equation}
w=\pi n-\frac{J(T)}{2\lambda}+{\cal O}(\frac{1}{\lambda^2}),
\end{equation}
where $n$ is an integer. To satisfy the boundary condition in Eq.~\eqref{bc:first}, we need $n=0$ and $J(0)=0$. To satisfy Eq.~\eqref{bc:second}, we must have
\begin{equation}
w(T_f,\lambda)=-\frac{J(T_f)}{2\lambda}=-f(T_f),
\end{equation}
which in turn requires $J(T_f)=f(T_f)=0$. This already negates the possibility of performing any nontrivial $\left(f(T_f)\ne0\right)$ $z$-rotation. However even error-free trivial rotations are impossible when the final condition, Eq.~\eqref{bc:zeroth}, is taken into account since this immediately gives $T_f=0$. We conclude that it is impossible to design a $J(t)$ that cancels field gradient errors to all orders. In reaching this conclusion, we have implicitly assumed that $J(t)$ is bounded and $T_f<\infty$.

\subsection{Finite rise time}
In this section we demonstrate that realistic pulses with finite turn-on/off time (rise time) would not substantially alter the \textsc{supcode} pulses. We focus on the general case of $h_0\neq 0$. In this case it is necessary to solve the Schr\"odinger equation numerically in order to find the rotation caused by a given pulse, and we denote the uncorrected rotation $\check{R}(\hat{r};\phi)$, defined
\begin{equation}
\check{R}(h_0 \hat{x} + J\hat{z};\phi)=\mathcal{T}\exp\biggl[-i \int_0^T \Bigl(\frac{h}{2} \sigma_x + \frac{J f(t,T)}{2} \sigma_z\Bigr)\,\mathrm{d}t\biggr]\quad,
\end{equation}
where $\mathcal{T}$ denotes time-ordering and the pulse shape is specified by a function $f(t,T)$. The pulse is of total duration $T$, found via numerical search such that the total rotation angle of $\check{R}$ is equal to $\phi$. In other words, to zeroth order in~$\delta h$, $\check{R}(\hat{r};\phi)$ is a rotation by an angle $\phi$ about an axis that deviates slightly from $\hat{r}$ due to the finite rise time. Using this uncorrected rotation $\check{R}$ we can construct a corrected $x$-rotation $\check{R}\check{I}$ with the same form as for the zero-rise time situation
\begin{equation}
    R(\hat{x};\phi)+ \mathcal{O}(\delta h)^2 = \tilde{R}(\hat{x};\phi)\check{I}_1(1/2,b'_1)^2\check{I}_2(1/2,b'_2,15) ,
\end{equation}
except that we now implement the $30\pi$ rotation in $\check{I}_2$ as two back-to-back $15\pi$ rotations 
\begin{align}
\check{I}_2(1/2,b'_2,15)=\tilde{R}(\hat{x};\pi)\check{R}(\hat{x}+b'_2 \hat{z};15\pi)^2\tilde{R}(\hat{x};\pi),
\end{align}
 because the finite rise time prevents performing one-pulse rotations of $2n\pi$, for $n$ integer. We perform a numerical search for $b'_1$, $b'_2$ such that the first-order error in $\delta h$ is again canceled, using the values $b_1$, $b_2$ from the zero-rise time case as a starting point.

\begin{figure}[t]
\centering
\includegraphics[width=0.9\textwidth]{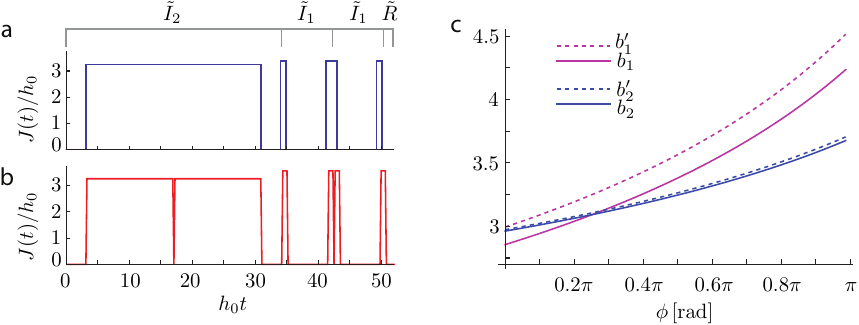}
\caption{{\bf {\footnotesize SUPCODE} pulse for rotation about the \boldmath $x$-axis by $\pi/2$ at $J_{\text{max}}/h_0\geq 5$.} (\textbf{a})~Example of a \textsc{supcode} pulse sequence, with instantaneous rise time (``boxcar pulse''). (\textbf{b}) ~The \textsc{supcode} pulse sequence with finite rise time, $\tau=0.0875/h_0$, corresponding to ${\sim} 1\,\text{ns}$ for $h_0\sim 50\,\text{neV}$. (\textbf{c})~Parameters for rotations about the $x$-axis for $0\le\phi\le\pi$. Solid lines are for boxcar pulses, dashed lines for finite rise time.}
\label{paraRx}
\end{figure}

\begin{figure}[t]
\centering
\includegraphics[width=0.9\textwidth]{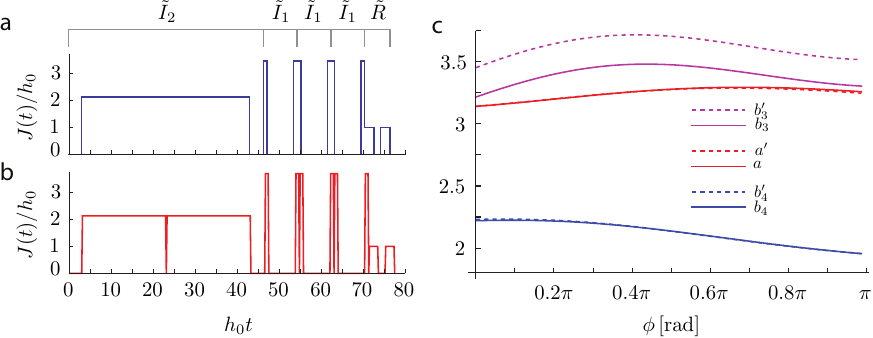}
\caption{{\bf {\footnotesize SUPCODE} pulse for rotation about the \boldmath $z$-axis by $\pi/2$ at $J_{\text{max}}/h_0\geq 5$.}  (\textbf{a})~Example of a \textsc{supcode} pulse sequence, with instantaneous rise time (``boxcar pulse''). (\textbf{b})~The \textsc{supcode} pulse sequence with finite rise time, $\tau=0.0875/h_0$, corresponding to ${\sim} 1\,\text{ns}$ for $h_0\sim 50\,\text{neV}$. (\textbf{c})~Parameters for rotations about the $z$-axis for $0\le\phi\le\pi$. Solid lines are for boxcar pulses, dashed lines for finite rise time.}
\label{paraRz}
\end{figure}

Our approach is valid for general pulse shapes, but as a simple example of such a pulse-shaping function $f(t,T)$ we consider trapezoidal pulses beginning and ending at $J=0$ and ramping up and down over a time~$\tau$:
\begin{equation}
f(t,T)=\begin{cases}
  t/\tau&0\le t\le\tau\\
  1&\tau<t\le T-\tau\\
  (T-t)/\tau&T-\tau<t \le T
\end{cases}\quad.
\end{equation}
As shown in Supplementary Figure~\ref{paraRx}, for physically reasonable values of $\tau$, the finite rise time causes only small perturbations to the pulse parameters.

For rotations about  $\hat{z}$ the only additional ingredient needed is when constructing the uncorrected $z$-rotation $\check{R}(h_0 \hat{x} + J_1\hat{z};\pi)\tilde{R}(\hat{x};\phi)\check{R}(h_0 \hat{x} + J_1\hat{z};\pi)$ it is necessary to search for $J_1 \approx h_0$ such that $\check{R}(h_0 \hat{x} + J_1\hat{z};\pi)$ generates a rotation around $\hat{x}+\hat{z}$. As shown in Supplementary Figure~\ref{paraRz}, the finite rise time again causes only small perturbations to the pulse parameters. On the scale of Fig.~4 of the main text, the error per gate is indistinguishable between the finite rise time and boxcar pulses.

In the preceding discussion, we assumed that the pulse profile~$f(t,T)$ is known precisely, allowing us to calculate numerically the necessary perturbation to the parameters of the corrected gate in order to compensate for the finite rise time. In an experimental implementation it should be possible to avoid the numerical calculation by performing the local search for the optimal gate parameters directly on the experiment, in which case it is not necessary to assume anything about~$f(t,T)$ apart from that it is fairly close to the ideal boxcar shape.

\end{document}